\documentstyle[psfig,emulateapj]{article}

\def\gta{{\gtrsim}}
\def\lta{{\lesssim}}

\def\be {{ \begin{equation} }}
\def\ee {{ \end{equation} }}
\def\fe {{\rm [Fe/H] }}

\begin{document}


\title{\bf Are Stars with Planets Polluted?} 
\author{N.  Murray\altaffilmark{1},
B. Chaboyer\altaffilmark{2} }

\altaffiltext{1}{Canadian Institute for Theoretical Astrophysics,
 60 St. George st., University of Toronto, Toronto, ONT M5S 3H8,
Canada; murray@cita.utoronto.ca}

\altaffiltext{2}{Department of Physics and  Astronomy, Dartmouth
College, 6127 Wilder Laboratory, Hanover, NH 03755-3528, USA;
chaboyer@heather.dartmouth.edu }

\begin{abstract}
We compare the metallicities of stars with radial velocity planets to
the metallicity of a sample of field dwarfs. We confirm recent work
indicating that the stars-with-planet sample as a whole is iron rich.
However, the lowest mass stars tend to be iron poor, with several
having $\fe <-0.2$, demonstrating that high metallicity is not
required for the formation of short period Jupiter-mass planets. We
show that the average [Fe/H] increases with increasing stellar mass
(for masses below $1.25M_\odot$) in both samples, but that the
increase is much more rapid in the stars-with-planet sample. The
variation of metallicity with stellar age also differs between the two
samples. We examine possible selection effects related to variations
in the sensitivity of radial velocity surveys with stellar mass and
metallicity, and identify a color cutoff ($B-V\gta0.48$) that
contributes to but does not explain the mass-metallicity trend in the
stars-with-planets sample. We use Monte Carlo models to show that
adding an average of $6.5M_\oplus$ of iron to each star can explain
both the mass-metallicity and the age-metallicity relations of the
stars-with-planets sample. However, for at least one star, HD 38529,
there is good evidence that the bulk metallicity is high. We conclude
that the observed metallicities and metallicity trends are the result
of the interaction of three effects; accretion of $\sim 6M_\oplus$ of
iron rich material, selection effects, and in some cases, high
intrinsic metallicity.

\end{abstract}


\section{Introduction}
Radial velocity surveys have established that $\sim5-7\%$ of
solar-type stars in the solar neighborhood have periodic velocity
variations with semiamplitude $K\gta 10\ {\rm m}/{\rm
s}$ (\cite{mcm}). The interpretation of these velocity variations as
being due to the presence of Jupiter-mass planets was clinched by the
observations of transits in HD 209458 (\cite{cha00,hen00}). The surveys
have revealed three surprising properties; first, many of the systems
have planets in extremely small ($0.03-0.1$ AU) orbits. Second, the
orbits, when not subject to tidal damping, are typically highly eccentric
($e\sim0.3$ being typical). Finally, the host stars are often highly
metal rich (\cite{gonzalez}).

In this paper we explore possible explanations for the high
metallicities of the stars-with-planets.  Two general classes of
explanation have been proposed for the high metallicities seen in
stars-with-planets; high intrinsic metallicities, and accretion of
metal rich material. A correlation between high intrinsic metallicity
and the presence of a radial velocity planet might arise if metal rich
gas disks are a prerequisite of either planet formation or planet
migration. Alternately, such a correlation would result from the
ingestion of rocky material or metal rich gas giant planets as a
result of the migration process.

Here we point out that a third explanation is currently viable, namely
selection effects. We discuss several possible selection effects.  We
point out an apparent color cutoff in the underlying samples producing
the radial velocity stars. We also discuss the bias associated with
the finite precision of the surveys combined with the apparent
increase in number of planets with decreasing planetary mass.

In this paper we examine all three types of explanation by comparing the
sample of radial velocity planet stars with a sample of dwarf stars in
the solar neighborhood recently studied by \cite{mcahn}. We find that
the available data is consistent with the notion that a substantial
amount of iron, of order six Earth masses ($6M_\oplus$) on average,
has been accreted onto the central stars in the radial velocity
systems. However, it is also possible, although in our judgment less
likely, that the trends in metallicity we see are due to the selection
effects mentioned above. The explanation based on high intrinsic
metallicity is inconsistent with the observation that several low
stellar mass radial velocity systems have substantially subsolar
metallicities; neither does it explain the variations in metallicity as a
function of stellar mass that we find. However, there is evidence that
{\em some} radial velocity systems are intrinsically metal rich; HD
38529 sits in the Hertzsprung gap, indicating that it currently
has a very deep convection zone, yet it has a high metallicity.

The paper is organized as follows. In \S \ref{data} we examine the
$B-V$ colors as a function of stellar mass, and the distribution of
metallicity as a function of stellar mass and age for both
stars-with-planets and the Murray et al. sample. We show that the
radial velocity sample has $B-V>0.48$, but in \S \ref{selection} we
argue that neither this nor the difficulty of achieving high precision
radial velocities for metal poor or blue stars completely explains the
metallicity trends we find. Neither does the decreasing sensitivity of
radial velocity surveys with increasing stellar mass (with the
subsequent weakening of the absorption lines), combined with the
reduced frequency of more massive planets, appear to explain the sharp
increase in \fe with increasing stellar mass. In \S \ref{pollution} we
use Monte Carlo models of stellar pollution to estimate the amount of
accreted iron needed to reproduce the observed metallicity trends,
assuming that they are not due to selection effects. In \S
\ref{discussion} we compare our results with recent results by other
authors, and we summarize our conclusions in \S \ref{conclusions}.

\section{Stellar Metallicity}\label{data}
We find the masses and ages of the known planet-bearing stars using
their HIPPARCOS parallaxes, combined with their $V$ magnitudes, their
effective temperatures $T_{eff}$ if available (or $B-V$ colors when
$T_{eff}$ is not known), and their metallicities \fe. For the bulk of
the stars with planets we use spectroscopically determined values of
the metallicity taken from the literature
\cite{castro,gv,g98,g99,gl,fischer,sim00,gltr}, but in the case of a
few of the more recently discovered systems we use color-based
metallicities, primarily from the CORAVEL web site
(http://obswww.unige.ch/~udry/planet/planet.html).  We fit to the grid
of stellar models described in \cite{mcahn}, to which we refer the
interested reader for further details.

In finding the ages and masses of the stars we have used models which
have uniform compositions. In the light of the findings presented
below, such models are only good as a lowest order approximation. We
have modified Chaboyer's stellar evolution code to handle polluted
models, but defer discussion of the results to later publications.

\begin{figure}
\plotone{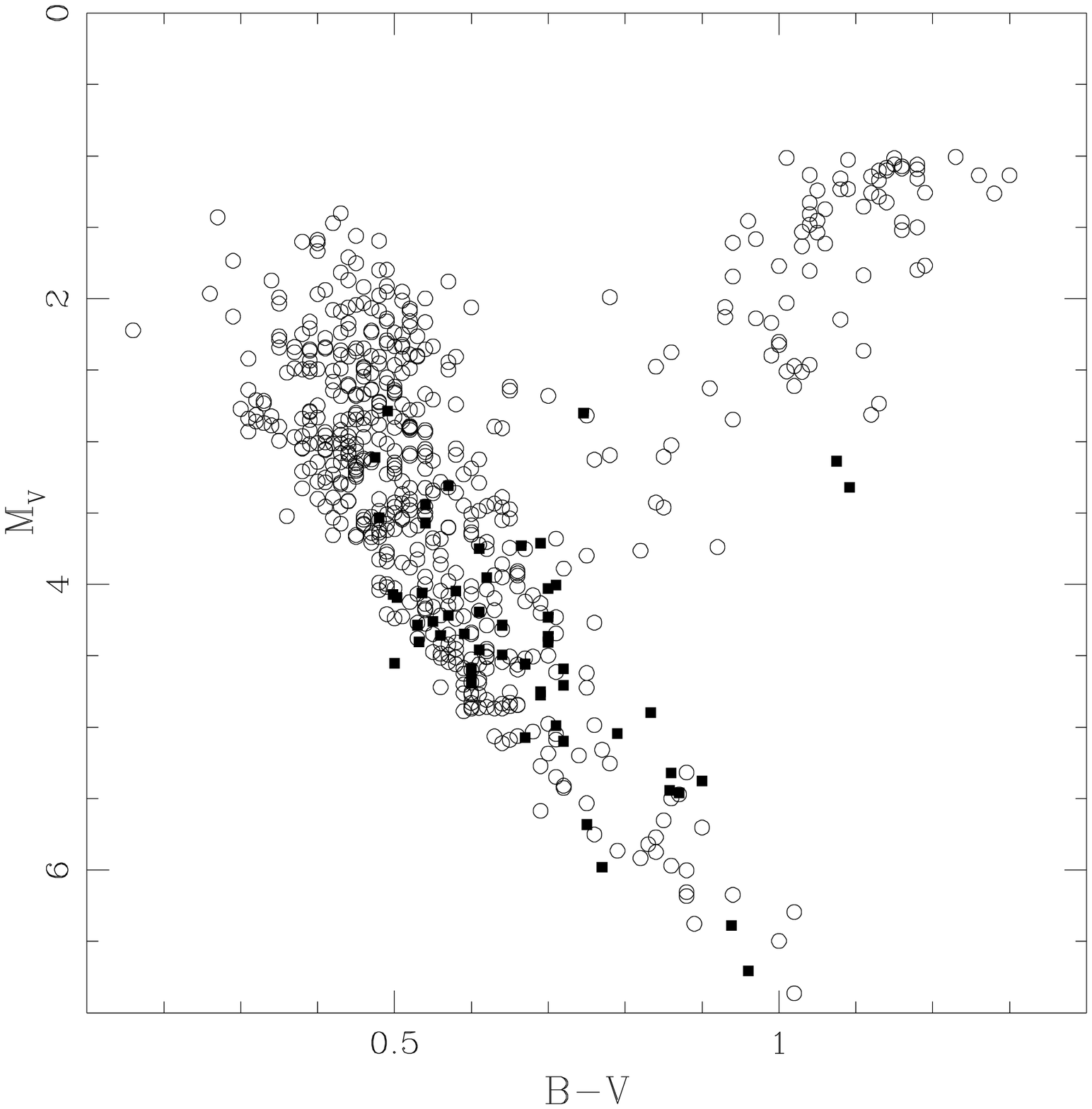}
\caption[HR diagram]{Color-magnitude diagram for both
stars-with-planets (filled squares) and the Murray et al. (2001)
sample (open circles). The Hertzsprung gap lies between the main
sequence to the lower left and the giant branch at the upper right;
only one star-with-planet lies in the gap, while 19 stars from the
Murray et al. sample are located there.
\label{Fig_HR}
}
\end{figure}

We compare the sample of stars-with-planets to the sample of 466 main
sequence stars, and to a sample of 19 slightly evolved (or
``Hertzsprung gap'') stars described in \cite{mcahn}. The
color-magnitude diagram for both samples is shown in Figure
\ref{Fig_HR}. The Hertzsprung gap stars are located in the sparsely
sampled region between the main sequence (running from the lower right
to the upper left hand corner of the plot) and the giants (in the uppr
right hand corner); the gap stars are just coming off the main
sequence, and have convection zones ten times more massive than the
surface mixing layer the star possessed while on the main sequence
(either convectively or rotationally mixed). They are useful as a
control, since their deep convection zones tend to minimize the effect
of any accretion of iron rich material that may have occurred while
the star was on the main sequence. 

We note that there are two giants stars in the current sample of
stars-with-planets. One of these, HD 177830, has a radius of about
$10R_\odot$ according to our models. However, its planet orbits at
$1.1$ AU, still at many tens of stellar radii.  Direct comparison
between the metallicities of dwarfs and giants are problematic, so the
giant stars are not shown in any of the following plots.

\begin{figure}
\plotone{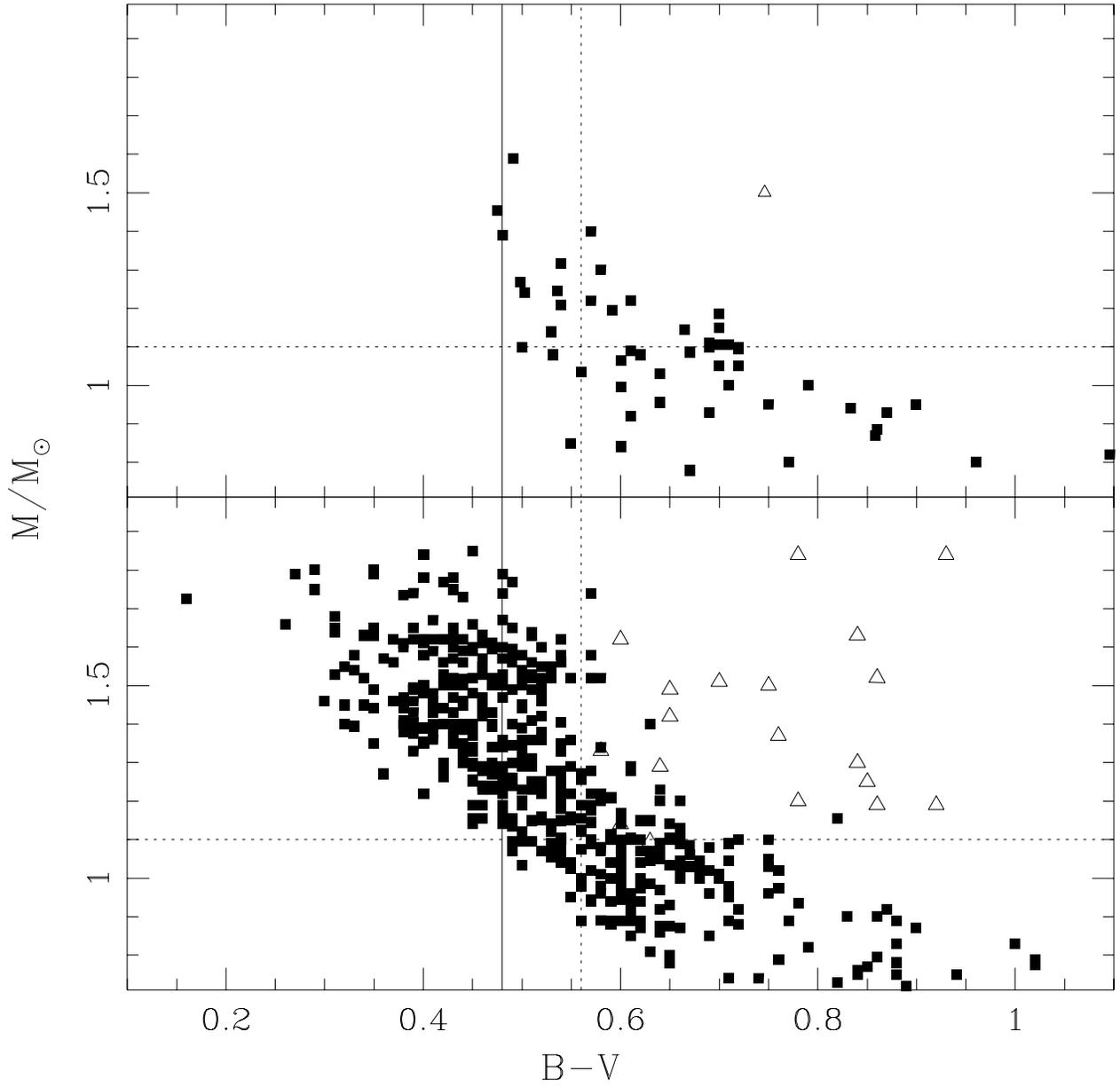}
\caption[Colors]{Stellar mass plotted as a function of stellar color
$B-V$ . The lower panel shows the sample of Murray et al., consisting
of 466 main sequence stars (solid squares) and 19 slighly evolved
'Hertzsprung gap' stars (open triangles). The upper panel shows 49
main sequence stars with planets (solid squares), and one slightly
evolved star with planets (open triangle). The dotted horizontal line
is at $M/M_\odot=1.1$. The solid vertical line is at $B-V=0.48$; we
attribute the lack of stars with planets to the left of this line to
an observational selection effect. This selection does not affect the
stars-with-planets sample below $1.1M_\odot$. The dotted vertical line
at $B-V=0.56$ corresponds to the cut off in the Santos et al. (2001)
sample of stars with no radial velocity planets.
\label{Fig_B_minus_V}
}
\end{figure}

Figure \ref{Fig_B_minus_V} shows the B-V color as a function of
stellar mass for both samples. The lower panel is the data for the
Murray et al. stars, the upper for stars with planets; in both panels
the solid squares are stars on the main sequence, while the open
triangles are stars in the Hertzsprung gap. Comparing the two panels
shows that the radial velocity surveys have a selection effect: with
two exceptions all the stars with planets have $B-V>0.48$ (the
horizontal solid line in the figure). In contrast, the Murray et
al. stars less massive than $1.5M_\odot$ range down to
$B-V\approx0.3$.

At fixed stellar mass, metal poor stars have lower values of $B-V$
than metal rich stars of the same age. Alternately, at fixed $B-V$,
increasing stellar mass implies increasing metallicity.  Thus the
current radial velocity surveys are biased against low metallicity
stars, and a plot of \fe versus mass based on such a selection
criterion will have a lower limit corresponding to increasing \fe
with increasing mass\footnote{The cutoff in $B-V$ could be a real
effect, as opposed to a selection effect. However, we note below that
the sample from which the planet bearing stars found by Mayor and
co-workers were drawn appears to have $B-V>0.56$}.

We note however that for stars less massive than $\sim 1.1M_\odot$
none of the stars in the Murray et al. sample have $B-V<0.48$, so for
stars less massive than this the color cutoff introduces no
bias. Furthermore, we show below that applying the $B-V$ cutoff to the
Murray et al. sample introduces only a slight change in an \fe vs
mass plot, a change not large enough to explain the difference between
the Murray et al. sample and the sample of stars with planets.

\begin{figure}
\plotone{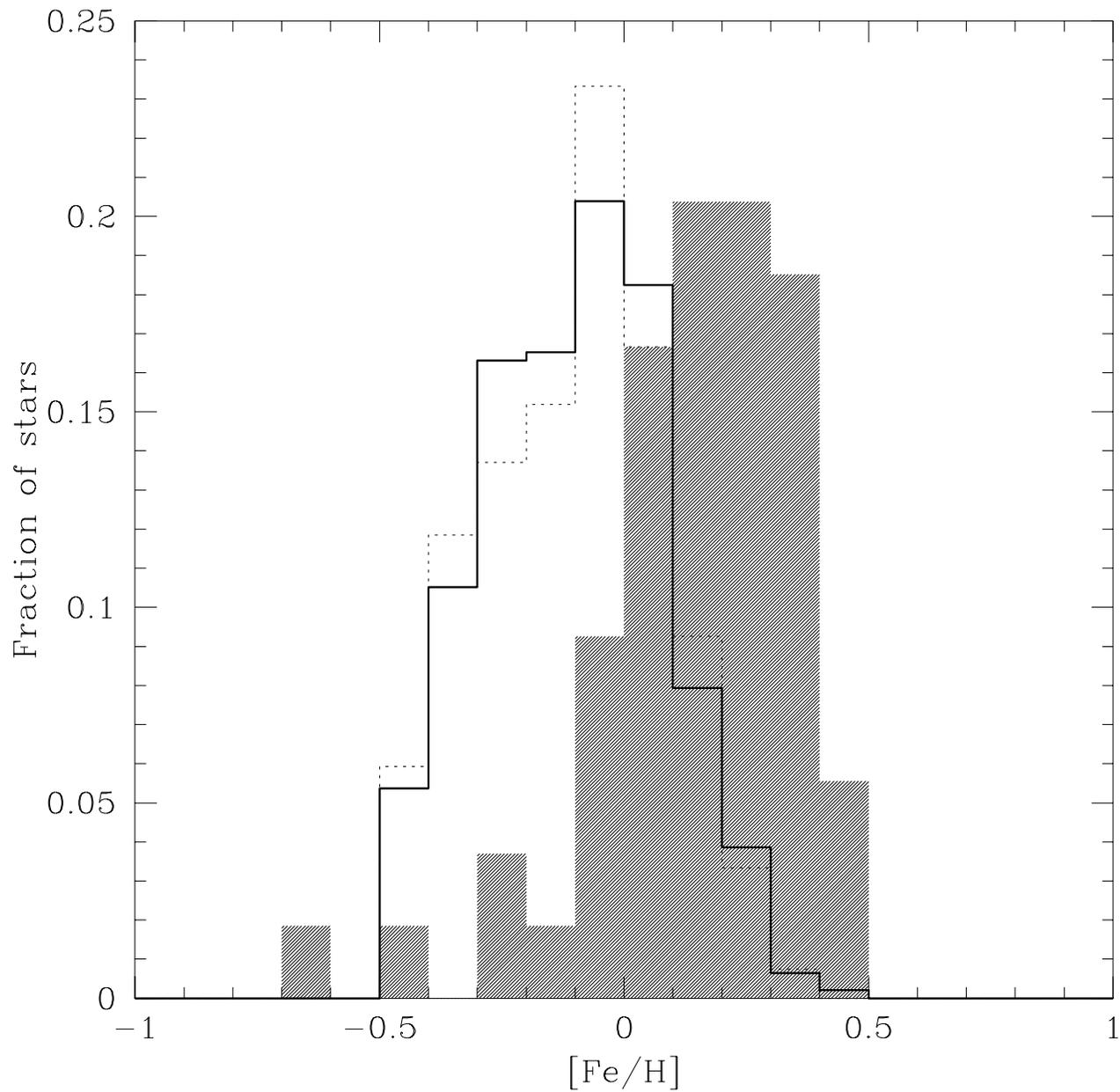}
\caption[Metallicity histogram]{Metallicity distribution of
stars-with-planets (shaded histogram) compared with the field dwarfs
in the Murray et al. (2001) sample. The dotted histogram is the result
of restricting the field dwarfs to have $B-V>0.48$, as appears to be
the case for the stars-with-planets sample (see Figure
\ref{Fig_B_minus_V}).
\label{Fig_Histogram}}
\end{figure}

Figure \ref{Fig_Histogram} shows the metallicity histograms of the
Murray et al. sample and the stars with planets. The unshaded
histogram drawn with the heavy solid line is for all the stars in the
Murray et al. sample, while the histogram drawn with the dotted line
includes only stars in that sample with $B-V>0.48$; the difference is
negligible. It is clear that the stars-with-planets sample is metal
rich relative to the other two samples, supporting the results of
earlier workers (\cite{sim00,laughlin,gltr}). It is also clear that this
difference is not due to the color selection criterion. 

A histogram of those stars-with-planets having $M_*\le 1.1M_\odot$ has
a lower average metallicity, but so does the analogous histogram for
the Murray et al. sample; the result is that there is still a
significant difference between the two samples. Since stars with
$M_*\le1.1M_\odot$ do not appear to be as blue as $B-V=0.48$, this
demonstrates again that the difference in \fe between the two
samples is not due to the apparent $B-V$ selection.

The excess metallicity of the stars with planets is even more striking
when the metallicity is plotted as a function of stellar mass, as in
Figure \ref{Fig_fe_mass} (see also Laughlin 2000). The lower panel
shows the metallicity of the stars from Murray et al. having
$B-V>0.48$.  The upper panel shows the metallicity of the stars with
planets (excluding two giants).  From this figure it is clear that,
while low mass stars-with-planets cover the same range of
metallicities as the Murray et al. sample, the high mass
stars-with-planets are all metal rich.

In the upper panel of Figure \ref{Fig_fe_mass} we have depicted those
stars having planets with semimajor axes $a>0.1$ AU by open
squares. The filled squares denote those stars with 'hot Jupiters',
having $a\le0.1$ AU. In Figure \ref{Fig_fe_mass_cold}a we show all
stars having planets with $a>0.5$ AU. It demonstrates that the
metallicity trends seen in the stars-with-planets sample (Figure
\ref{Fig_fe_mass}) are not strong functions of the semimajor axis of
the planets. We discuss the significance of this result in \S
\ref{discussion}. 

The solid line in both panels shows the boundary between Chaboyer's stellar
models with $B-V>0.48$ (up and to the left of the line) and those with
$B-V<0.48$ below and to the right). The calculated values of $B-V$
depend on the age of the model; both pre-main sequence and evolved
stars are redder than stars of the same mass and metallicity on the
main sequence. In plotting the solid line we have taken models on the
main sequence at or near the age at which $B-V$ is minimized. If all
the stars in both samples were near the zero age main sequence,
essentially all would be above and to the left of this solid line
(assuming our colors are good matches to those of real stars). This is
true for the stars-with-planets sample, but not for the Murray et
al. sample. Our models indicate that the latter sample has many stars
which are evolving off the main sequence, and which are currently
redder than $B-V=0.48$ even though younger stars with their mass and
metallicity would have $B-V<0.48$. The location of the Hertzsprung gap
stars (plotted as open triangles) is consistent with this
interpretation. The width of the main sequence in Figure \ref{Fig_HR}
near the upper end ($M_V\approx 2$) is also consistent with this picture.

If the trend of increasing \fe with increasing mass were due to the
(apparent) color selection of the radial velocity surveys, we would
expect that the bulk of the metallicities of the parent stars would be
above and to the left of this solid line, with the possibility that
some slightly evolved stars would be below and to the right, as is
seen in the Murray et al. sample. In fact all the stars with planets
{\em are} above the $B-V=0.48$ line; even for stars with mass below
$1.1M_\odot$, when the color selection should not matter (according to
Figure \ref{Fig_B_minus_V}) the stars-with-planets are all well above
the line. We conclude that the mass-metallicity trend seen at the high
mass end of the stars-with-planets sample is not due entirely to the
$B-V$ cutoff; as noted above, in the case of stars with
$M_*<1.1M_\odot$, the color cutoff has no effect.

Binning the data by mass (Figure \ref{Fig_fe_mass_binned}) again shows
that the metallicity properties of the Murray et al. sample differ
from those of the stars with planets. The data points in this figure
show the average \fe in mass bins $0.1M_\odot$ wide, along with the
associated standard errors (variance divided by the square root of the
number of star less one). The open squares correspond to the Murray et
al. sample, the filled squares to a subset of that sample with
$B-V>0.48$. As noted above, the color selection criterion does not
effect the metallicity distribution for stars below $\sim 1.1M_\odot$;
even for more massive stars, the differences between the average
\fe with and without applying the color cutoff, while noticeable, are
much smaller than the differences between the Murray et al. sample and
the stars with planets sample (filled triangles). 

We note that the average metallicity of the stars with planets rises
much more rapidly with increasing mass than either the Hertzsprung gap
or the main sequence stars without planets. This  strongly suggests
that the stars with planets have accreted iron rich material after they
reached the main sequence. We quantify this below.

It is also instructive to plot metallicity as a function of stellar
age, since younger stars are expected to be metal rich when compared
to older stars. As seen in the lower panel of Figure
\ref{Fig_fe_age_binned}, there is a correlation between stellar age
and metallicity in the Murray et al. sample; younger stars are, on
average, more metal rich than older stars. The average value of \fe
rises steadily from $\sim-0.2$ at 10 Gyrs to $\sim+0.05$ at the
present. Compared to this rather regular rise, the age-metallicity
distribution of the stars-with-planets sample (the upper panel) is
odd. It jumps rather abruptly from $\sim0.0$ (or solar) at $\sim10$
Gyrs to $+0.2$ for stars of all ages less than $\sim 5$Gyrs.

We must account for the difference in $B-V$ between the two samples to
make a fair comparison; we do so by plotting only those stars with
$B-V>0.48$ and mass less than $1.45M_\odot$ as open squares in the
bottom panel. There is a slight difference, but the result is nothing
like the data in the upper panel.

\subsection{IS THE RISE OF [FE/H] WITH MASS A SELECTION
EFFECT?}\label{selection} 
The sample of stars-with-planets was apparently drawn from a stellar
sample with a selection criterion of $B-V\gta0.48$. This by itself
cannot explain the the rapid rise of metallicity with stellar mass, as
demonstrated by Figures \ref{Fig_fe_mass} and
\ref{Fig_fe_mass_binned}.

However, there is another possible selection effect. The radial
velocity technique relies on a cross-correlation between a known
spectrum (usually provided by an iodine cell placed in the beam) and
the stellar spectrum. Stars that are more massive and hence hotter
have weaker lines than less massive stars; similarly metal poor stars
have weaker lines than metal rich stars. Thus massive metal poor stars
will have weak, sparse spectra, and achieving high precision radial
velocities for such stars is likely to be problematic.

This line of argument suggests that part of the metallicity trend we
see might be due to the inability of radial velocity surveys to
identify planets around massive metal poor stars. To evaluate this
properly one could examine the radial velocity errors as a function of
stellar mass and metallicity, or as a function of $B-V$, since
increasing mass and decreasing metallicity both tend to reduce
$B-V$. We urge observers to do so. 

Lacking this information, we instead examine the distribution of the
semiamplitude $K$ with stellar mass; see Figure
\ref{Fig_K_mass}. There does appear to be a mass-dependent lower
envelope; for stars more massive than $\sim1.1M_\odot$, the minimum
$K$ increases with increasing stellar mass. We also note that for
stars less massive than the sun, the minimum $K$ increases with {\em
decreasing} stellar mass. Both of these trends may be due simply to
the small numbers involved; most stars with planets have very nearly
solar masses, with only a few stars more massive than $1.2M_\odot$ or
less massive than $0.85M_\odot$. However, it is possible that the
apparent trends are due to a $K$ selection effect. Here we will assume
that it is, and show that such a selection does not appear to explain
the rapid rise in \fe with stellar mass.

To do so we plot \fe versus stellar mass, but only for those
systems having $K>50\ {\rm m}/{\rm s}$ (see Figure
\ref{Fig_fe_mass_cold}b), on the assumption that the current surveys
are reasonably complete for $K$ this large. The trend with stellar
mass is unchanged.

There is a third possible selection effect related to the masses of
the parent stars. The distribution of planet masses appears to
indicate that there are more low mass planets than high mass
planets. A power law provides a rough fit, $N(m)\sim m^{-\alpha}$,
with $\alpha$ of order 1. Suppose that the current radial velocity
surveys can detect $K=K_{min}$ for all stars up to
$1.6M_\odot$. A star of $1.6M_\odot$ must have a planet twice as
massive as a star of $0.8M_\odot$ in order to produce a wobble of
amplitude $K_{min}$. The more massive star is then only $2^{-\alpha}$
as likely to have such a planet, assuming stars of different masses
have the same chance of having a planet of a fixed mass. Examining
equal numbers of high and low mass stars would then yield more planets
around low mass stars. If the sample is not complete near $K_{min}$,
the high mass stars would then tend to be more metal rich than the low
mass stars, since the low metallicity high mass stars with low $K$
would be harder to detect. We believe however, that the current
samples are reasonably complete for $K>50\ {\rm m}/{\rm s}$, so this
seems unlikely to explain the strong trend seen in the \fe versus
stellar mass plot.

We tentatively conclude that the trend of increasing \fe with
increasing stellar mass is a real physical effect. The odd
distribution of \fe with stellar age for the stars-with-planets sample
is also hard to explain as the result of a selection effect. Assuming
that stars-with-planets only form in high metallicity clouds also
fails to explain the age-metallicity relation, since it would predict
uniformly high metallicities. In the next section we present Monte
Carlo models of polluted stellar populations and compare them to the
data.

\section{MODELING THE POLLUTION}\label{pollution}
To model the pollution we follow Murray et al. (2001). The outer
layers of the star are assumed to be well mixed down to a depth which
depends on the mass and bulk metallicity of the star. For stars less
massive than $1.2M_\odot$, the mass of this mixed layer is
assumed to be given by the mass of the surface convection zone. For
more massive stars Murray et al. used the observed lithium abundances
of open clusters to infer the mass of the surface mixing
layer. Briefly, the abundance of lithium as a function of stellar mass
(or effective temperature) shows a dramatic dip around $1.4M_\odot$
\cite{bt86}. The dip is believed to be due to the thermonuclear
destruction of lithium at depths where the temperature exceeds $\sim
3\times 10^6$K. This strongly suggests that surface material is mixed
down into the star well below the bottom of the convection zone, which
is very thin in a $1.4M_\odot$ star. Murray et al. describe a crude
empirical model for the depth of this surface mixing layer as a
function of stellar mass and composition; we use that model in this
section.

The stars in the Murray et al. sample show a distinct jump in \fe
around $1.5M_\odot$, just above the lithium dip (the open squares in
Figure \ref{Fig_fe_mass_binned}). They argue that the stars in their
sample, the bulk of which are not known to have Jupiter-mass planets,
have accreted iron rich material after having reached the main
sequence. Using their model for the depth of the mixing layer, they
compare Monte Carlo simulations of polluted stellar populations to the
observed mass-metallicity distribution to infer that, on average, the
stars in their sample have accreted $\sim 0.4M_\oplus$ of iron. (We
reproduce the result in Figure \ref{Fig_theory}.)

The Hertzsprung gap stars afford a test of this conclusion. If most
stars are polluted, then when they evolve and develop or deepen
surface convection zones, their metal rich surface layers will be
mixed with the relatively metal poor inner layers, reducing the
surface iron abundance.  As can be seen from Figure \ref{Fig_theory},
the Hertzsprung gap stars (shown as open triangles) are metal poor
compared to unevolved stars of the same mass.

Figure \ref{Fig_theory} also shows that a polluted model can fit all
three data sets (including the stars-with-planets sample) with only
one free parameter, the average mass of accreted material.  The
unpolluted model is shown as the dotted line, which passes through the
error bars associated with the Hertzsprung gap stars. As mentioned
above, the solid line, which is a best fit model for the unevolved
stars in the Murray et al. sample, corresponds to an average mass of
accreted iron of $0.4M_\oplus$. The dashed line running through the
stars-with-planets data corresponds to a model with an average of
$6.5M_\oplus$ of accreted iron. This plot makes a clear prediction;
stars above $\sim1.5M_\odot$ that have short to moderate period
Jupiter-mass planets will have very high metallicities. The plot
suggests \fe of order $0.5$.

We caution that this prediction assumes that our unpolluted models are
good representations of stars that have surface layers much more metal
rich than their interiors (by about $0.5-0.7$ dex). This assumption
must be checked using polluted models. Nevertheless, we feel that
stars with short period Jupiters having masses larger than 
$\sim1.5M_\odot$ will have metallicities \fe substantially above $0.2$
dex.

The same Monte Carlo models yield the histogram of polluted stars
shown in Figure \ref{Fig_theory_histogram}. The model is shown by the
open histogram, on which is superimposed the histogram of
stars-with-planets. In both samples we include only stars less massive
than $1.45M_\odot$, since there are only two stars in the
stars-with-planets sample more massive than this value. The agreement
is excellent.

Finally, the Monte Carlo models give the run of average \fe with
age. The solid lines in Figure \ref{Fig_fe_age_binned} give the
results. We see that the model with $0.4M_\oplus$ of added iron gives
a good fit to the Murray et al. sample, while the model with
$6.5M_\oplus$ of added iron fits the stars-with-planets data. We
stress that the underlying (unpolluted) variation of metallicity with
stellar age is the same in both Monte Carlo models. Thus pollution can
explain the variation of \fe with mass, and the variation of \fe
with stellar age, in both samples.

This result bolsters  our confidence that the stars-with-planet sample
has suffered a substantial amount of pollution.

\section{DISCUSSION}\label{discussion}
\begin{figure}
\plotone{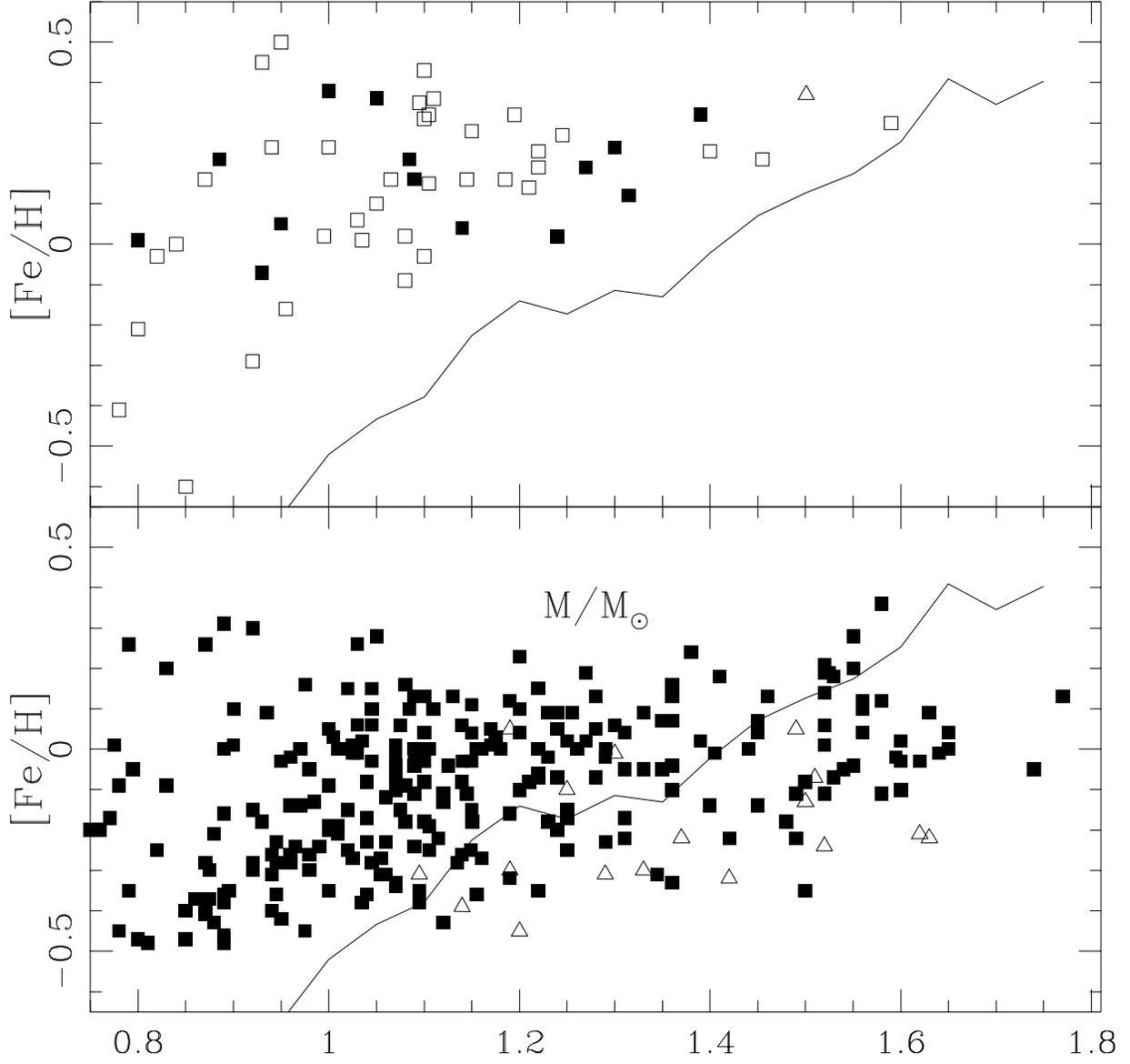}
\caption[Metallicity versus stellar mass]{Stellar metallicity as a
function of stellar mass, where the mass is obtained by fitting to our
stellar models. The upper panel is for stars with planets, the lower
for stars in the Murray et al. sample. Filled squares represent
unevolved stars having planets with $a\le0.1$ AU, open squares
unevolved stars having planets with $a>0.1$ AU, while open triangles
represent Hertzsprung gap stars. The solid line in both panels
corresponds to the mass and metallicity of main sequence stars with
$B-V=0.48$, calculated at the age where $B-V$ is minimised. Stars
below and to the right of this line can have $B-V>0.48$ if they are
either very young or slightly evolved, hence the overabundance of
Hertzsprung gap stars in this region. The fact that the stars with
planets lie well above this line suggests that their high metallicity
is not simply due to the $B-V$ selection. 
\label{Fig_fe_mass}}
\end{figure}

\begin{figure}
\plottwo{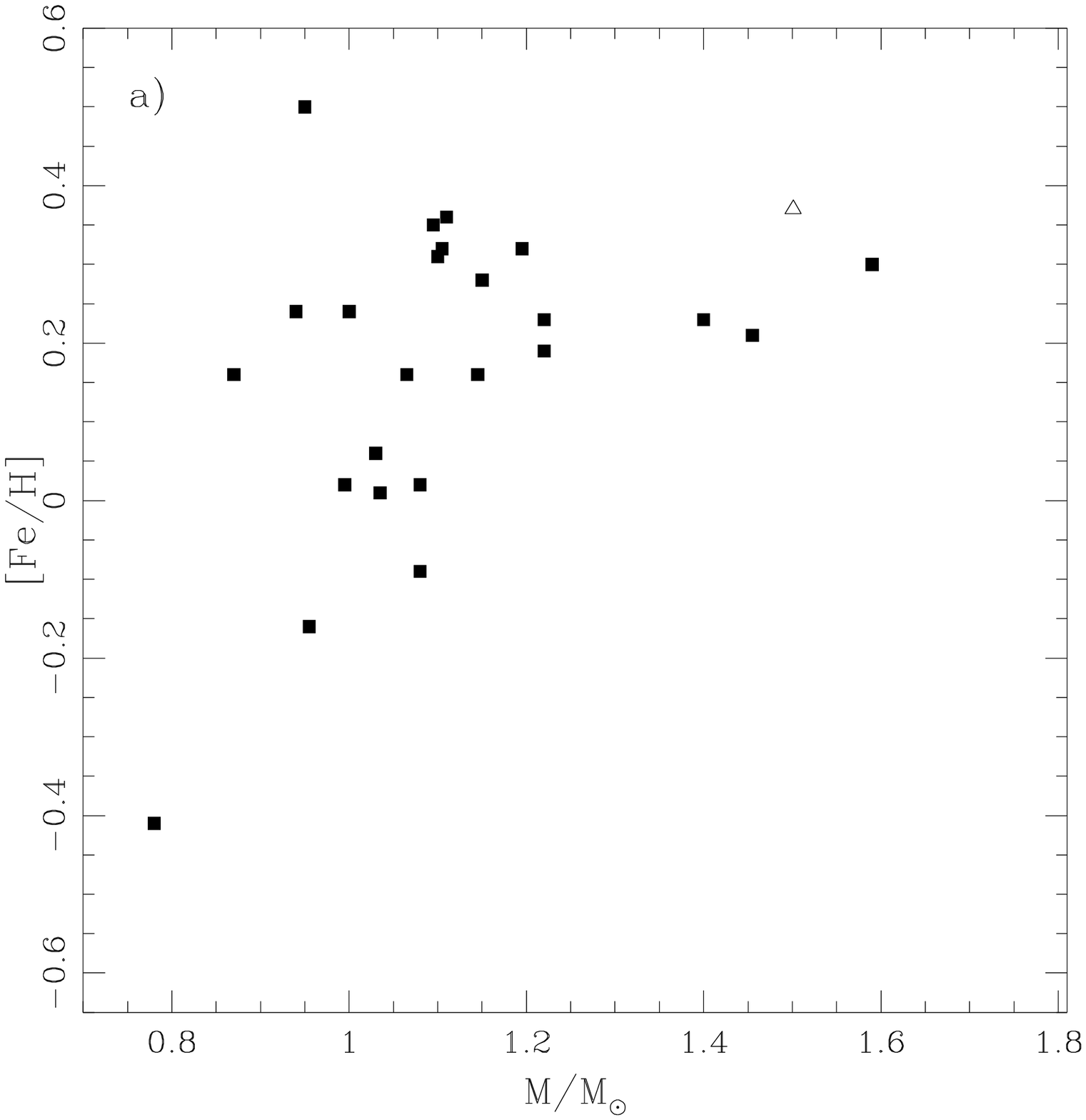}{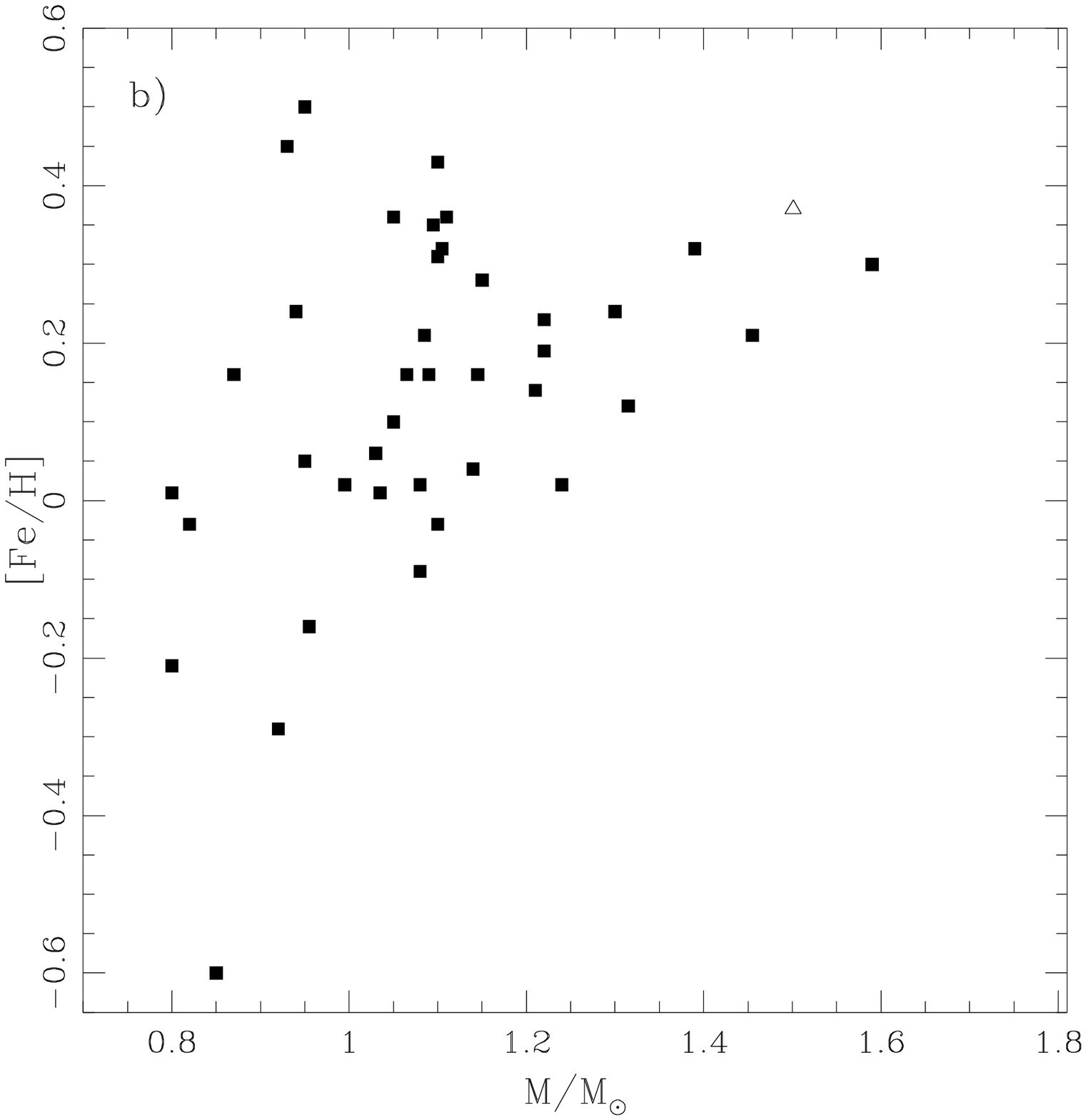}
\caption[Metallicity versus stellar mass]{a) As in the upper panel of
Figure \ref{Fig_fe_mass}, but only for stars whose innermost planet
has $a>0.5$ AU. b) The same, but for stars with $K>50\ {\rm m}/{\rm s}$.
\label{Fig_fe_mass_cold}}
\end{figure}

\begin{figure}
\plotone{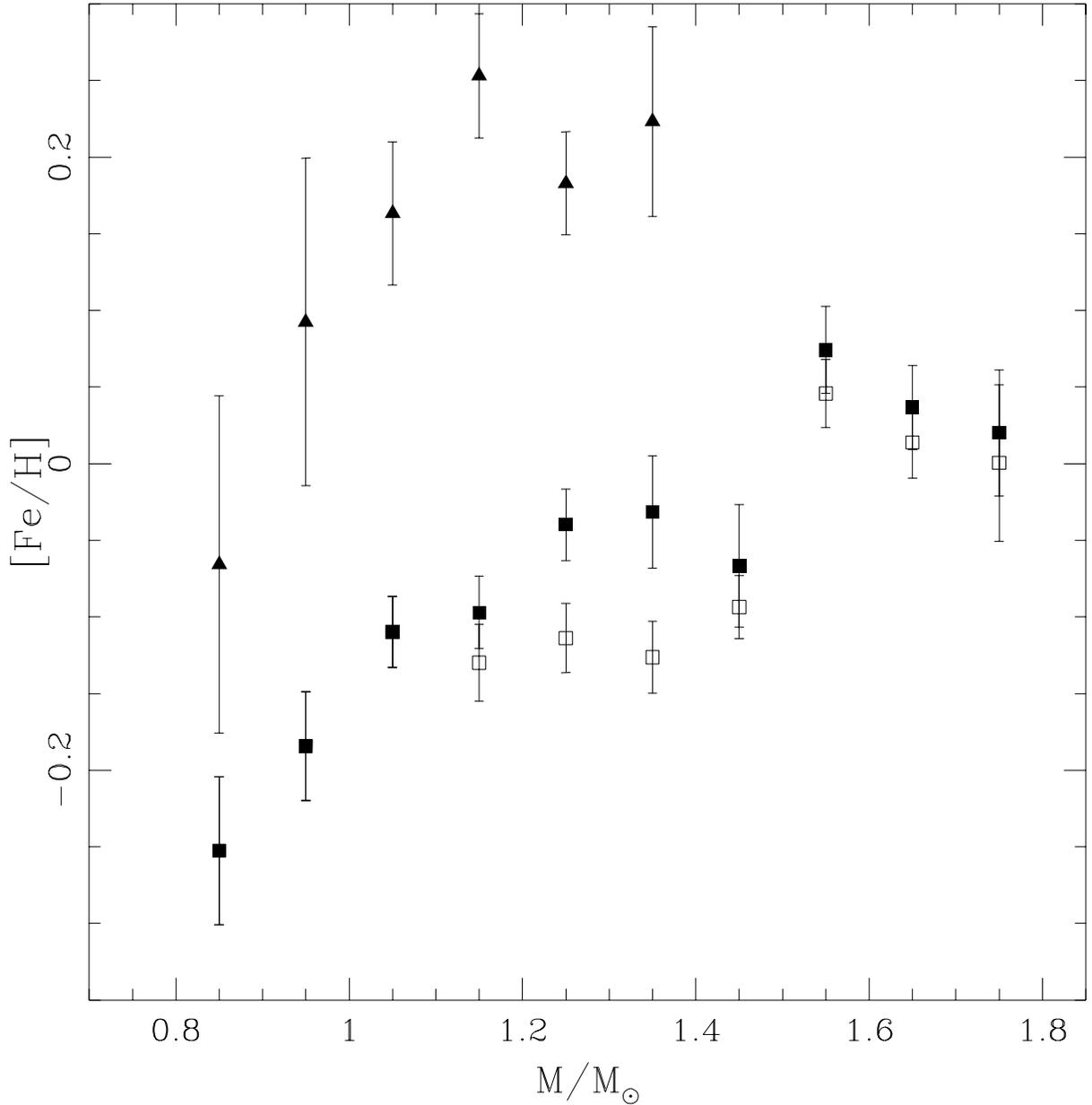}
\caption[Metallicity versus stellar mass, binned]{Stellar metallicity
binned as a function of stellar mass. Open squares represent unevolved
stars from the Murray et al. sample, while filled squares result from
requiring $B-V>0.48$. As expected, there is no difference between the
open and filled squares below $1.1M_\odot$, so the open squares cannot
be seen below that mass. The filled triangles represent the stars with
planets. Note that even for stars with masses $\lta 1.1M_\odot$,
where the color selection plays no role, the stars with planets have
much higher metallicity than the stars in the Murray et
al. sample., and that the metallicity increases rapidly with stellar
mass (more rapidly than in the Murray et al. sample). 
\label{Fig_fe_mass_binned}}
\end{figure}

\begin{figure}
\plotone{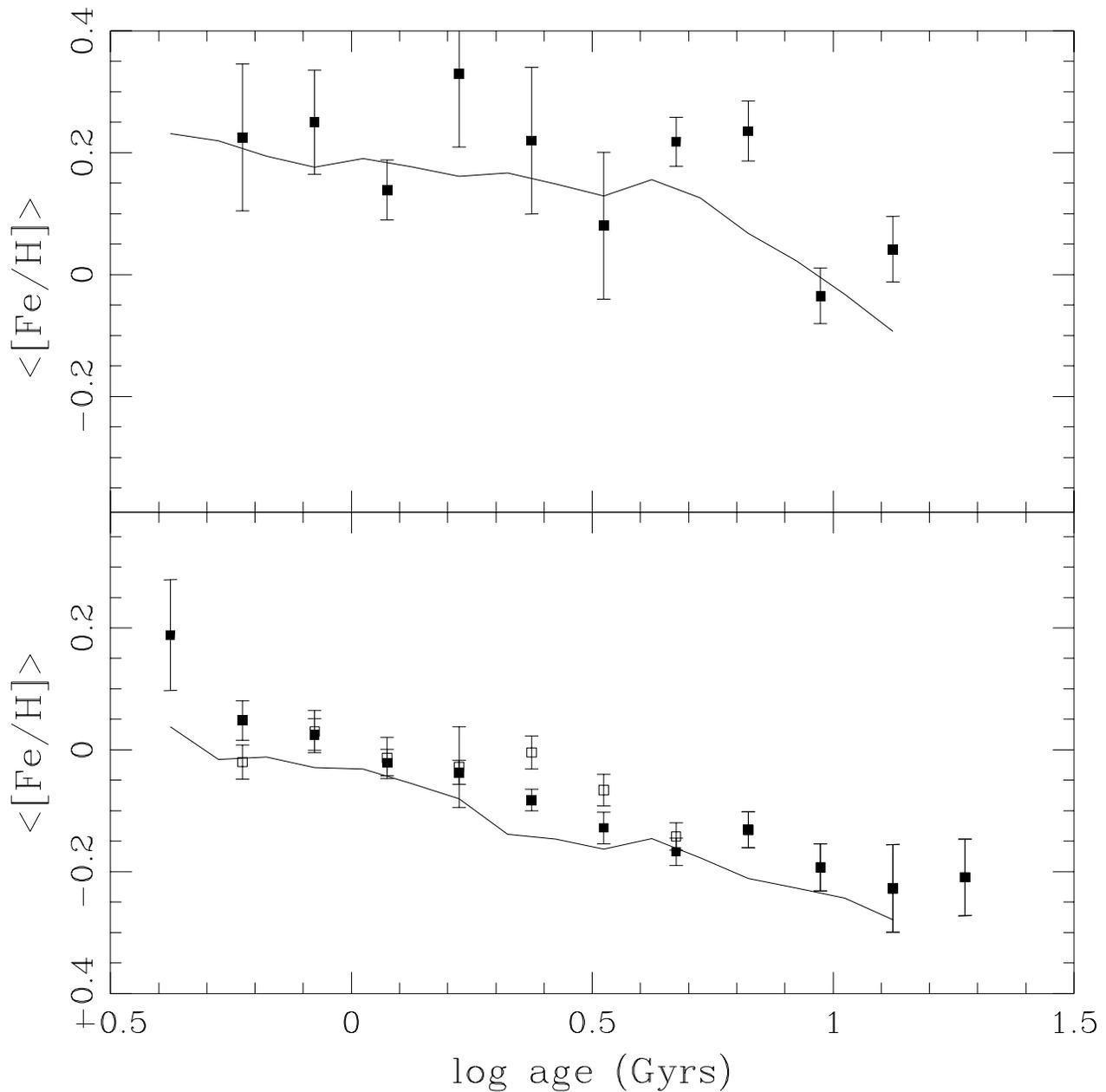}
\caption[Metallicity versus age, binned by age]{Average stellar
metallicity in age bins of width $\Delta\log(age)=0.15$. There are roughly 20
stars per bin for the Murray et al sample (lower panel), but only a
few per bin for the stars with planets sample (upper
panel). Nevertheless, it is clear that the two distributions differ
dramatically. The solid lines are the predicted average \fe values
for polluted models; in the lower panel the average accreted iron mass is
$0.4M_\oplus$, while in the upper panel it is $6.5M_\oplus$.
\label{Fig_fe_age_binned}}
\end{figure}

\begin{figure}
\plotone{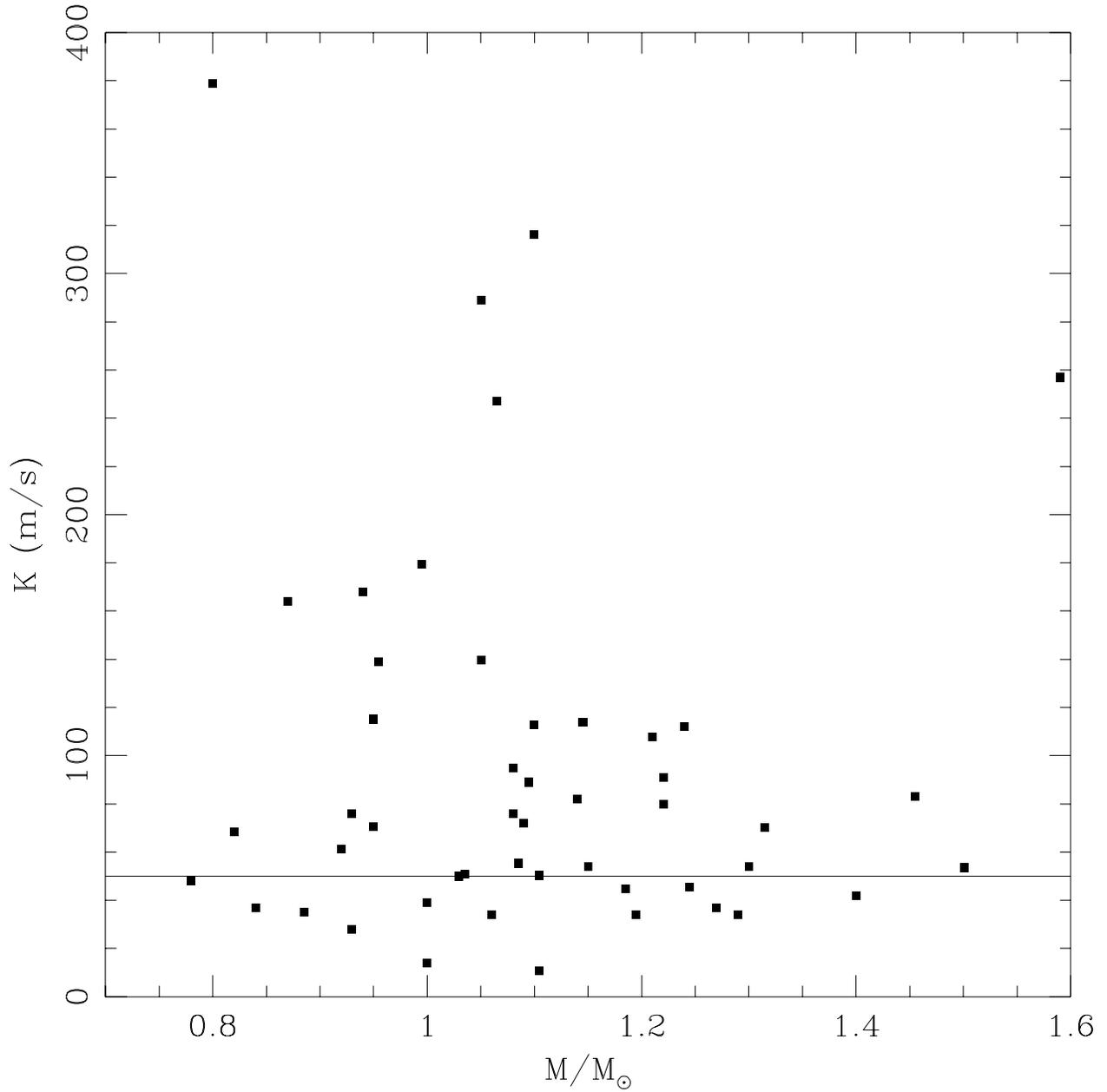}
\caption[K versus mass]{The amplitude $K$ of the radial velocity
plotted against stellar mass for the known extrasolar planets. There
is a hint of a lower envelope for high mass stars, running from 
$10\ {\rm m}/{\rm s}$ at $1.1M_\oplus$ to $\sim 50\ {\rm m}/{\rm s}$
at $1.5M_\oplus$. The current radial velocity surveys may well be able to
detect planets around $1.5M_\odot$ stars at lower radial velocities,
but we take $50\ {\rm m}/{\rm s}$ as a conservative estimate for the
radial velocity  at which the  surveys are complete in the mass
range $0.8\le M/M_\odot\le 1.5$. Note that there are several systems
with $K>400$ which are not shown in this plot. Note also that there is
a hint of a selection against small $K$ in stars less massive than
$1M_\odot$. 
\label{Fig_K_mass}}
\end{figure}

\begin{figure}
\plotone{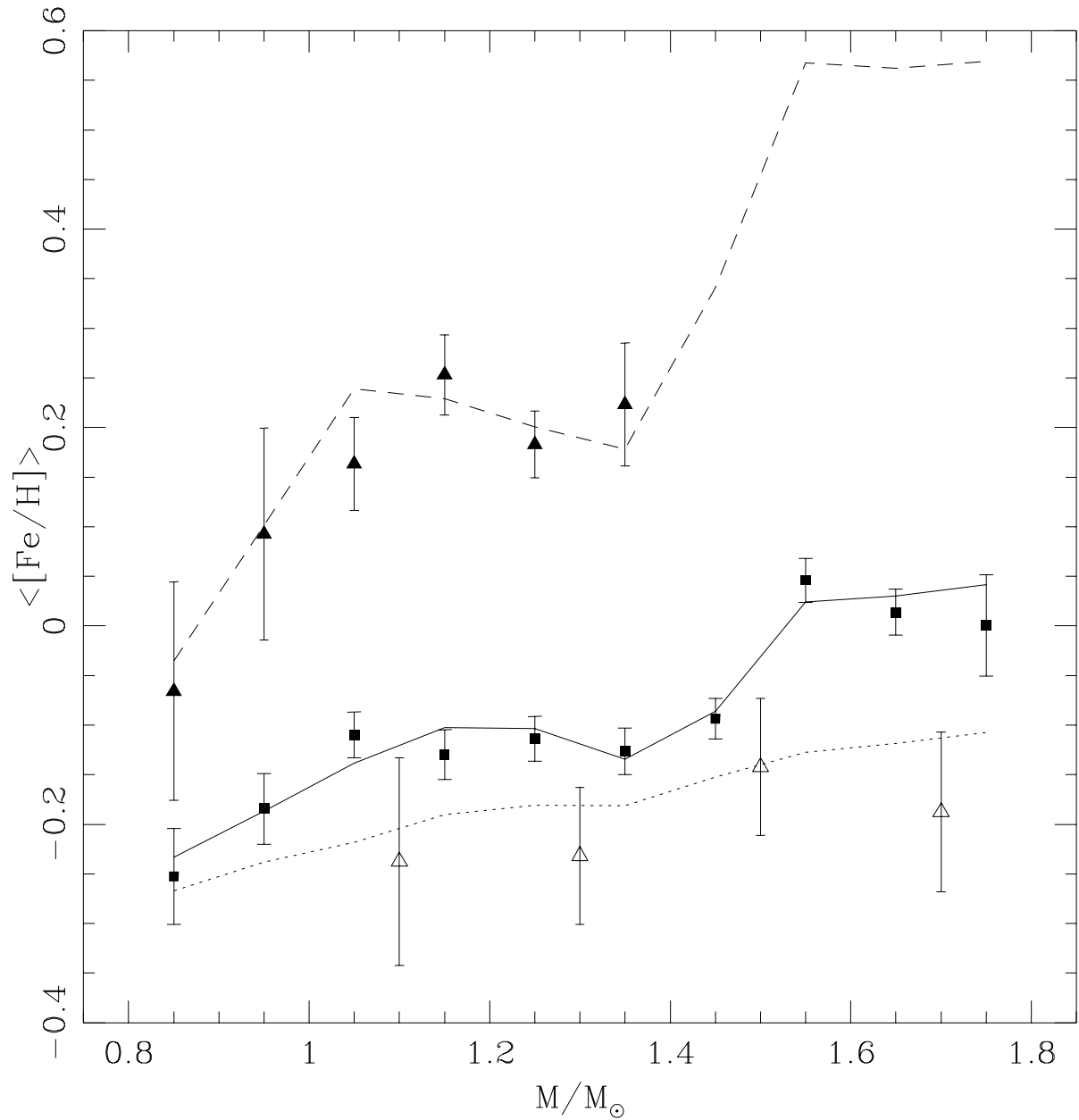}
\caption[metallicity versus mass]{ The average metallicity of three
populations of Monte Carlo polluted stars compared with the three
different observed distributions. The solid curve is a model in which
stars accrete a gaussian distributed amount of iron with mean
$0.4M_\oplus$. It is a best fit model for the Murray et al. (2001)
sample, shown as the solid squares. No cutoff in $B-V$ has been
applied to the model or the data. The dotted line is the same model,
except that the surface mixing region is assumed to have deepend by a
factor of $10$ (in mass); it gives an acceptable fit to the
Hertzsprung gap stars. The dashed curve is a model in which stars have
accreted $6.5M_\oplus$ of iron on average. It gives an acceptable fit to
the sample of stars with planets.  
\label{Fig_theory}}
\end{figure}

\begin{figure}
\plotone{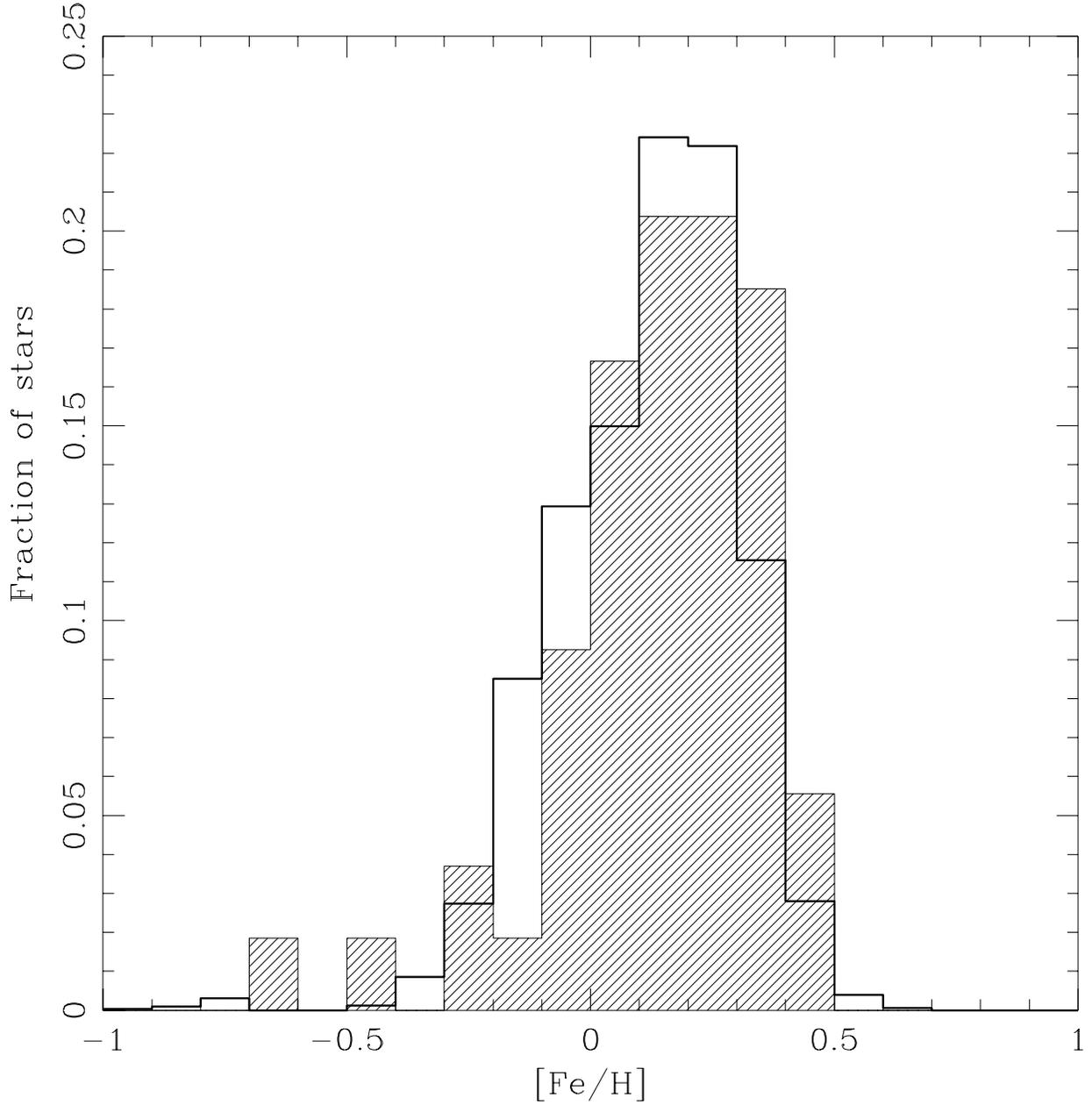}
\caption[Histogram of Fe/H]{A comparison between the \fe histogram of
known stars with planets and the result of a Monte Carlo experiment in
which iron is added to the envelopes of a population of unpolluted
stars. The unpolluted population is that described in
\cite{mcahn}. The mean added iron mass is found by optimizing the fit
of $\fe$ as a function of stellar mass (see Figure
\ref{Fig_theory}) and is $6.5M_\oplus$. We  assume
variance of $1/3$ the mean, or $2.2M_\oplus$. 
\label{Fig_theory_histogram}}
\end{figure}

Four recent papers \cite{laughlin,sim00,sim01,gltr} have presented
figures similar to the upper panel of our Figure
\ref{Fig_fe_mass}. Laughlin also shows data for photometrically
determined \fe as a function of stellar mass, much like the bottom
panel of Figure \ref{Fig_fe_mass}. Both Laughlin and Gonzalez et
al. conclude that the data are consistent with pollution of the outer
envelopes of the stars-with-planets by iron rich material, {\em after
the stars have reached the main sequence}.

In contrast, both Santos et al. papers conclude that the data ``support the idea
that a star needs to be formed out of a metal rich cloud to form giant
planets''. They argue that their results rule out pollution.

In view of the fact that there are now seven stars-with-planets having
\fe$<-0.05$, and that four of these have \fe$<-0.2$, the
statement that a star needs to be born in a metal rich cloud to form
giant planets seems unwarranted. 

From Figure \ref{Fig_fe_mass} we see that the distribution of
metallicity for stars with $0.75\lta M_*\lta 0.95$ is nearly the same
for stars-with-planets and stars without (known) planets.  From Figure
\ref{Fig_fe_mass_binned} we see that the average metallicity rises
with mass for both classes of stars, although it rises much more
rapidly for stars-with-planets. In contrast, it rises only very slowly
for Hertzsprung gap stars. Murray et al. argue that the rise seen in
their sample arises partly from pollution, and partly from the fact
that less massive stars tend to be older than more massive stars (the
`age-mass relationship'), and older stars tend to be metal poor.

We note that the age-mass relationship we find for the
stars-with-planets sample is very similar to that of the Murray et
al. sample, yet the \fe versus stellar mass relation for the
stars-with-planets sample rises much more rapidly than that of the
Murray et al. sample. We have shown above that accretion of
$6.5M_\oplus$ of iron will produce such a rise. In the absence of a
known selection effect capable of producing such a dramatic trend,
this suggests that the stars-with-planets have accreted substantial
amounts of iron rich material.

\cite{sim01} also show, in their Figure 4, a histogram of
stars-with-planets together with sample in which they added
$15M_\oplus$ of iron to their sample of 43 stars with low limits on
$K$ (that is, stars which don't have Jupiter-mass planets in short to
moderate period orbits). They restricted their stars to have mass less
than $1.2M_\odot$. They find that there are too many 'polluted' stars
at high values of \fe compared to the stars-with-planets
sample, and that the steep fall at the high metallicity end of the
distribution is not reproduced. They conclude that that a simple
pollution model cannot explain the observations.

Our results, illustrated in Figures \ref{Fig_theory_histogram} and
\ref{Fig_theory}, contradict with this statement. The main differences
appear to be that Santos et al. added $15M_\oplus$ of iron, about
three times the amount added in our best fit model, and that they took
a smaller mass (equal to the mass of the convection zone) for the
surface mixing layer of even their most massive stars. Since the iron
content of the convection zone of their $1.2M_\oplus$ star of solar
metallicity is about $1.4M_\oplus$, they find stars with \fe
approaching unity when they add $15M_\oplus$ of iron. Our models use a
slightly more massive surface mixing layer for stars above
$1.2M_\odot$ (taken from an empirical fit to lithium abundance data,
see Murray et al. 2001). More importantly, we add less iron, so we do
not find many stars with metallicities above $0.5$ dex.

Santos et al. (2001) also plot \fe versus mass (their Figure 5) for
their sample of planet-less dwarfs. The latter sample shows a lower
envelope of increasing \fe with increasing mass, which tracks the
lower envelope of the stars-with-planets. In contrast, as can be seen
in our Figure \ref{Fig_fe_mass}, while we see a lower envelope that
increases with increasing mass in our sample, it has a much lower
value of \fe at any given stellar mass. Clearly their sample of dwarfs
differs from the one we employ.

In fact a quick check shows that all but two stars in the \cite{sim01}
planet-less sample satisfy $B-V> 0.56$. The bluest star has $B-V=0.51$;
it appears as an outlier in the lower right hand corner of their
Figure 5. Had they allowed for planet-less stars as blue as $0.48$ (as
found in the stars-with-planets sample, and applied to the stars in
the Murray et al. sample as plotted in Figure (\ref{Fig_fe_mass})
they would have seen many planet-less stars at every mass having \fe
much lower than the stars-with-planets sample.

The peculiar behavior of \fe with stellar age for the
stars-with-planets sample is a second indication that pollution has
occurred in these stars. Restricting $B-V$ to be larger than $0.48$ (as
shown by the open squares in the bottom panel of Figure
\ref{Fig_fe_age_binned}) or even $0.55$ in the Murray et al. sample
does not produce the very sharp rise at ages around 10 Gyrs. However,
the polluted model that fits the \fe versus mass data naturally
produces a good fit to the \fe versus stellar age data.

As pointed out in Murray et al. (2001), slightly evolved (or
Hertzsprung gap) stars offer a possible test of the pollution
scenario; they should have, on average, lower metallicities if most
stars are polluted. Our fits to evolutionary tracks show that only one
star in the current stars-with-planets sample, HD 38529, is such a
star; this star can be seen in the Hertzsprung gap in Figure
(\ref{Fig_HR}). HD 38529 has a rather high \fe$=0.37$. We conclude
that at least some of the high metallicities seen in this sample are
likely to be due to a high primordial metallicity. We stress however,
that high primordial metallicities are not required to produce radial
velocity planets, nor will they produce the steep trend in \fe seen
in the low mass stars. More stars-with-planets will be discovered in
the Hertzsprung gap as the radial velocity surveys proceed, which will
allow an unbiased test of the pollution scenario.

\subsection{Giant Planet Accretion, or Planetesimal Accretion?}
Laughlin and Adams (1997) showed that accretion of Jupiter-mass
planets pushed into their parent stars during the lifetime of the
protoplanetary disk would produce little or no pollution of low mass
stars. However, Lin (1997) has speculated that Jupiter-mass bodies may be
accreted after the gas disk has dissipated. We argue that such a
scenario is unlikely to explain the difference between the
stars-with-planets sample and the Murray et al. sample, assuming the
steep increase in \fe in the former is not a selection effect.

Figure \ref{Fig_fe_mass_cold} demonstrates that the elevated
metallicities seen in the stars-with-planets does not require the
presence of a very short period Jupiter; 9 of the systems with
\fe$>0.2$ exhibit innermost planets having semimajor axes exceeding
1 AU, and several have $a>2$AU. HD 27442 is a solar mass star
($M_*=1.06M_\odot$) with a planet of mass $M\sin i=1.42M_J$ at $1.18$
AU on a nearly circular orbit; the star has \fe$=0.2$. If this star
ingested a Jupiter mass planet, it did so with out the aid of the
remaining planet.

Figure \ref{Fig_fe_mass} shows that there is a distinct shortage of
low metallicity stars-with-planets having masses greater than about
$0.9M_\odot$. If this is due to pollution, then essentially every
star-with-planet is polluted; if a significant fraction were not
polluted, then we would see some low \fe but slightly evolved high
mass stars-with-planets, as well as unevolved low mass
stars-with-planets closer to the $B-V$ cutoff. 

Our models show that the average amount of accreted iron in the
stars-with-planets sample is of order $6M_\oplus$. In contrast, no
more than about $10\%$ of the the stars in the Murray et al. sample
could have accreted that much material, or the distribution of \fe
would be double peaked, contradicting the observations.

The tight correlation between the presence of a radial velocity planet
and the accretion of several Earth masses of iron is difficult to
explain as the result of the accretion of a Jupiter-mass body. It
requires that most or all stars with radial velocity planets accrete
Jupiter-mass bodies (after $10-20$ Myrs) independent of the semimajor
axis of the remaining planet(s), while less than $20\%$ of stars
lacking radial velocity planets do so. (Note that the rather weak
metallicity mass trend found by Murray et al. shows that no more than
$10-20\%$ of stars lacking radial velocity planets could have accreted
a Jupiter-mass planet with the metallicity of Jupiter.)

It is difficult to believe that the presence of a Jupiter-mass planet
at $1-3$ AU, as seen in several systems, can ensure the accretion of a
second Jupiter-mass planet. We note that our own Jupiter, at $5.2$ AU,
almost certainly did not cause the accretion of any Jupiter-mass
objects onto our sun 20 Myrs after it formed. 

On the other hand Jupiter did cause the accretion of a substantial
mass of asteroids onto the sun, of order $2-3M_\oplus$. It is likely
that Jupiter migrated inward $\sim0.1$ AU during the ejection of
interplanetary material, but the presence of the inner asteroid belt,
as well as the regular orbits of both Jupiter and Saturn, suggest that
Jupiter did not migrate several AU. 

The small orbits of the currently known extrasolar planets are
consistent with the notion that they migrated inward several AU. If
there was a massive planetesimal disk inside the original location of
these planets, a substantial fraction, of order $5-10\%$, of the
material in the disk would have been dropped on the star
(\cite{1998Sci...279...69M,quillen,mph}) as a result of resonant
interactions between the planet and the planetesimals. If the
migration proceeds to very small semimajor axes ($\lta0.2$ AU) a
second process becomes efficient, namely scattering of planetesimals
by the planet onto the star. This will add another $\sim5-10\%$ of the
disk mass onto the star \cite{1998Sci...279...69M,hansen}.

If the planetesimal disk is {\em responsible} for the inward migration
(\cite{1998Sci...279...69M}), then the amount of accreted material
will be larger for more massive planets. This effect should be most
dramatic in more massive stars, since they have less massive surface
mixing regions. A plot of \fe versus $M\sin i/M_J$, for stars more
massive than $0.8M_\odot$ (Figure \ref{Fig_msini}) does hint at such a
trend of increasing metallicity with increasing mass, but the trend is
significant only at the two-sigma level. Increasing the size of the
stars-with-planets sample will allow this question to be answered more
definitively.

Another possible signature of planetesimal accretion is the relative
enhancement of volatile (e.g., C, N) versus refractory (Fe, Ni)
elements. Asteroids and terrestrial planets are depleted in noble
gases; terrestrial planets (but not most asteroids) are also strongly
depleted in elements such as carbon, presumably because the formed
from material that condensed at high temperature. By contrast, Jupiter
and the other giant planets in our solar system are carbon rich (they
are presumably also iron rich, but there is no direct evidence).
Thus, accreting material that condensed very close to the star would
enhance the iron abundance, but not the carbon abundance. Accreting
rocky material that formed farther from the star would enhance both
carbon and iron, as would the accretion of Jupiter.

We conclude that stars having radial velocity planets in orbits of
order $0.2$ AU or smaller are likely, in the planetesimal accretion
scenario, to have accreted material that is rich in refractories such
as iron and magnesium, but volatile poor, since the planetesimals in
such small orbits condense from the protoplanetary disk at high
temperature. There is some evidence indicating that this is the case
(\cite{smith}).

\section{CONCLUSIONS}\label{conclusions}
Previous work has shown that stars possessing radial velocity planets
appear, on average, to be metal rich compared to field stars of the
same mass and age (\cite{gonzalez,laughlin,sim00,gltr}). We confirm this
result. Our work has emphasized that part of this difference is due to
a color selection effect: the stars-with-planets appear
to be drawn from a sample chosen to have $B-V>0.48$ (or $0.56$ in the
case of the sample presented in Santos et al. 2001). As a result, the high
mass stars included in the present surveys {\em have} to have high
metallicity, skewing the distribution of \fe in the resulting
sample of stars-with-planets.

However, we have shown this color cutoff does not affect stars with
$M_*\le 1.1M_\odot$; even for the subset of stars less massive than
this, the stars-with-planets still have higher average metallicity
than our control sample. Applying the same $B-V$ cutoff to the control
sample does in fact increase the average \fe of the high mass
stars, but not by enough to explain the difference between the two
samples. We conclude that the high metallicity of the
stars-with-planet sample is not due to the color selection effect. Nor
is the steep rise in \fe as a function of stellar mass in the range
$0.8M_\odot<M_*<1.1M_\odot$, or the lack of a trend in \fe as a
function of stellar age, explained by the color selection effect.

Similarly, none of these observational results appears to be explained
by a $K$ selection effect, although this question clearly needs more
work. 

Our Monte Carlo simulations of pollution show that all three
observations (high average \fe, rapid increase in \fe with
stellar mass, and the lack of a trend in \fe with age) can be
produced by the addition of $\sim6M_\oplus$ of iron to those stars
which have radial velocity planets.

Laughlin (2000) has pointed out that measurements of \fe in those
stars-with-planets having stellar companions can be used to test the
pollution hypothesis. We point out another test, examining
stars-with-planets in the Hertzsprung gap. The one star in the current
sample that resides in the gap is metal rich, suggesting that its bulk
metallicity is high; no pollution is required for this object,
although it might have had an even higher (apparent) metallicity while
on the main sequence.

A third test will be provided by the discovery of planets around stars
more massive than $\sim1.6M_\odot$; if pollution is occurring on the scale
suggested here these stars should be very metal rich. Further
theoretical work is needed to determine exactly how metal rich they
should be.

\begin{figure}
\plotone{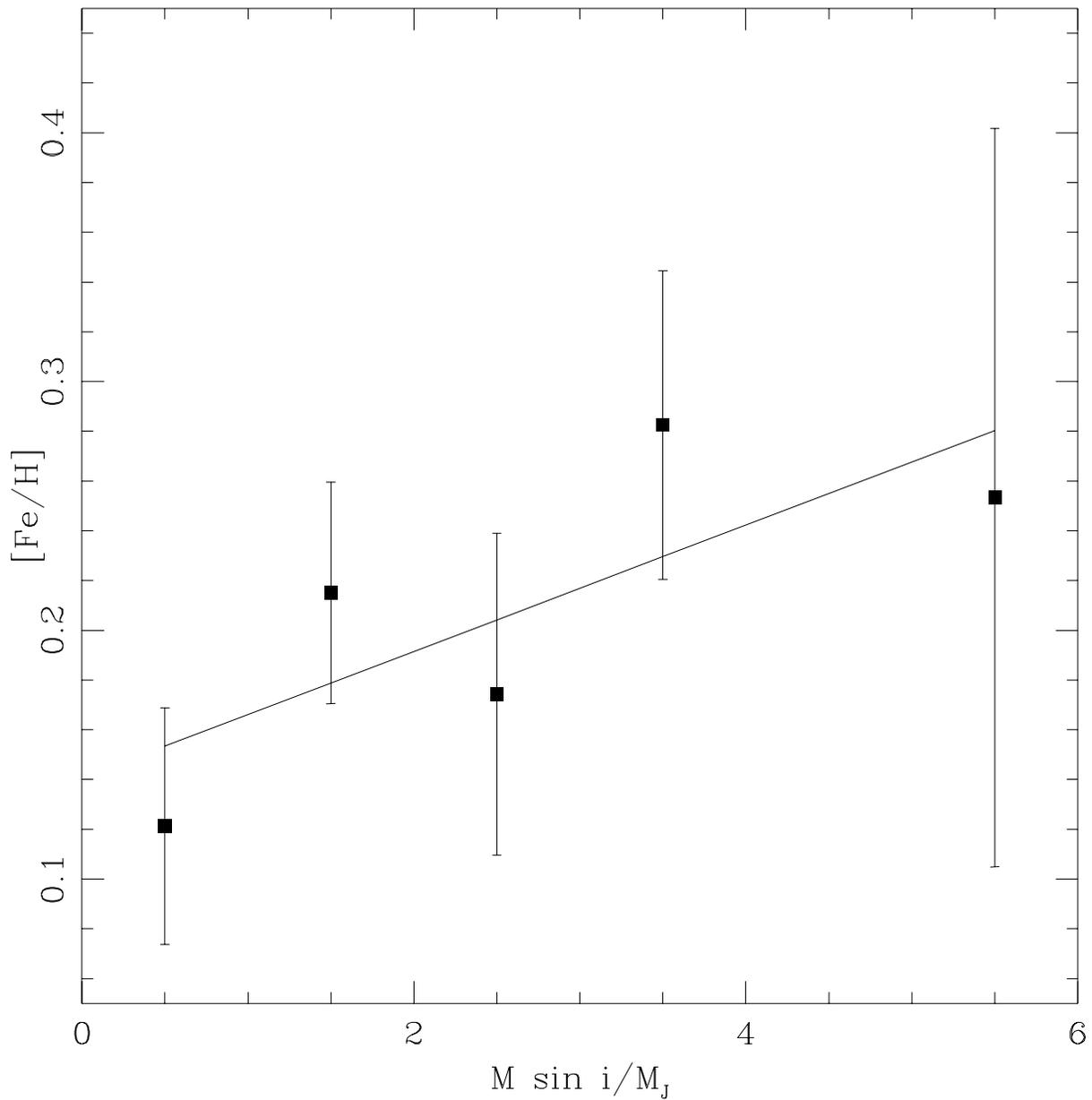}
\caption{The average metallicity of the stars-with-planets, binned by
$M\sin i$ of the planet. the stars are required to be more massive
than $0.80M_\odot$ since the convection zones of less massive stars
are to deep to show significant pollution. There is a hint of an
increase in \fe with increasing planetary mass. The straight line is a
least squares fit, $\fe=a(M\sin i/M)J)+b$, with $a=0.025\pm 0.012$.
\label{Fig_msini}}
\end{figure}

We argued that any accreted iron is unlikely to have been added
in the form of a gas giant planet, and suggested that trends in the
abundances of volatile versus refractory elements can be used to
distinguish between gas giant accretion, and planetesimal
accretion (\cite{smith}). A second trend, namely higher \fe in systems
with more massive planets, as hinted at in Figure \ref{Fig_msini},
could also be used to distinguish between planetesimal accretion and
gas giant accretion. Establishing the reality of either trend would
strongly suggest that pollution plays a significant role in such
systems. 

\acknowledgements Support for this work was provided by NSERC of
Canada, by NASA, and by a NSF CAREER grant to Brian Chaboyer. Dr
Chaboyer is a Cottrell Scholar of the Research Corporation. This
research made use of the SIMBAD database, operated at CDS, Strasbourg,
France.

\clearpage

\end{document}